**Discovery and Characterization of an Extremely Deep-Eclipsing Cataclysmic Variable: LSQ172554.8-643839**

**Short title: Discovery of LSQ172554.8-643839**

David Rabinowitz[1], Suzanne Tourtellotte[2], Patricio Rojo[4], Sergio Hoyer[4], Gaston Folatelli[4], Paolo Coppi[2], Charles Baltay[3], Charles Bailyn[2]

[1]Center for Astronomy and Astrophysics, Yale University, P.O. Box 208120, New Haven CT, 06520-8120 USA, david.rabinowitz@yale.edu
[2]Dept of Astronomy, Yale University, P.O. Box 208101, New Haven CT, 06520-8101 USA
[3]Dept of Physics, Yale University, P.O. Box 208120, New Haven CT, 06520-8120 USA
[4]Department of Astronomy, Universidad de Chile, Casilla 36-D, Santiago, Chile

## ABSTRACT

We report the discovery of an eclipsing cataclysmic variable with eclipse depths > 5.7 magnitudes, orbital period 94.657 min, and peak brightness V~ 18 at J2000 position $17^h 25^m 54^s.8$, $-64°38'39"$. Detected by visual inspection of images from Yale University's QUEST camera on the ESO 1.0-m Schmidt telescope at La Silla, we obtained light curves in B, V, R, I, z and J with SMARTS 1.3-m and 1.0-m telescopes at Cerro Tololo and spectra from 3500 to 9000 Å with the SOAR 4.3-m telescope at Cerro Pachon. The optical light curves show a deep, 5-min eclipse immediately followed by a shallow 38-min eclipse and then sinusoidal variation. No eclipses appear in J. During the deep eclipse we measure V-J > 7.1, corresponding to a spectral type M8 or later secondary, consistent with the dynamical constraints. The estimated distance is 150 psec. The spectra show strong Hydrogen emission lines, Doppler broadened by 600 - 1300 km s$^{-1}$, oscillating with radial velocity that peaks at mid deep eclipse with semi-amplitude 500 ± 22 km s$^{-1}$. We suggest that LSQ172554.8-643839 is a polar with a low-mass secondary viewed at high inclination. No known radio or x-ray source coincides with the new object's location.

INTRODUCTION

Cataclysmic variables (CVs) are binaries consisting of a white dwarf (WD) accreting material from a low-mass companion star, usually a red dwarf, orbiting closely enough to fill its Roche lobe (Warner 1995, hereafter W95, provides a comprehensive review). Orbital periods range from ~80 min to many hours, with a period gap from roughly 2 to 3 hours. Many CVs are extremely variable. Usually the infalling gas forms an accretion disk and a bright spot where the gas impacts the edge of the disk. The disk episodically brightens owing to accretion instability, outshining the WD and secondary by many orders of magnitude. For some CVs (polars) the WD magnetic field strength is so extreme it synchronizes the WD rotation to the orbit and disrupts the formation of an accretion disk. The infalling gas from the companion star then streams directly onto the surface of the WD, forming one or more bright emission regions near the WD surface. CVs have long been the target of photometric and spectroscopic time-series studies over wide wavelength ranges (X-ray to infrared) and timescales (seconds to years) to characterize the nature of their extreme variability, the disk-accretion phenomena, the effects of magnetic fields and gravitational radiation, and the influences of these phenomena on the CV evolution.
For eclipsing CVs, high time resolution light curves reveal the locations of the active emission regions, the relative masses of the WD and the secondary, and their radii relative to the size of the disc and the binary separation. Assuming a mass/radius relation for either the WD or the companion then fixes the absolute sizes and mass scales of the system (Wood 1986, Littlefair 2008).

Here we report discovery and follow-up observations of an unusual CV incidently detected by visual inspection of wide-field optical images taken with Yale University's 160-Megapixel QUEST camera (Baltay et al. 2007) on the 1.0-m Schmidt telescope of the European Southern Observatory (ESO) at La Silla, Chile. The QUEST camera is a mosaic of 112 CCDs covering ~10 square degrees. The La Silla – QUEST survey (LSQ) is an ongoing, automated search to discover nearby supernovae, distant solar-system objects, and other transients. Previously mounted on the 1.2-m Oschin Schmidt telescope at Palomar, we recently moved the camera to the ESO Schmidt where we are conducting a fully automated, all-sky survey in a custom wide-band filter (4000 to 7000 Å, here referred to as the "$Q_{st}$" band). The new variable was detected by visual inspection of three $Q_{st}$-band images taken at 12 minutes intervals on 2010 Apr 10 in order to provide follow up astrometry for an asteroid discovered with the WISE spacecraft. Blinking the images to detect the asteroid revealed the presence of a star ~100 times brighter than the detection limit, alternately appearing and disappearing. The object was noted independently by D. Rabinowitz and by Joseph Masiero of the WISE team at the Jet Propulsion Laboratory (personal communication, 2010 Apr 14).

Our initial follow-up observations revealed a periodic light curve with very deep (> 5 mag), short-duration (~5 min) eclipses and short period (P ~ 95 min). The only sources known to exhibit eclipses with comparable depth and frequency are CVs. Very few, however, are known to show such deep eclipses. Figure 1 shows the eclipse depth versus peak magnitude for the 160 eclipsing compact binaries with measured period and eclipse depth cataloged by Ritter and Kolb (2003). Of these, only 3 have depths greater than 5 mag. Two are polars (SDSS J015543.40+002807.2 and RXJ0501.7-0359, see Burwitz et al 1999 and O'Donoghue et al 2006) and the other is a detached binary (PG1550+131 = NN Ser, see Parsons et al 2010). Recognizing the unusual character of the light curve, we organized a modest follow-up campaign to verify the target as a CV, and to steer future follow-up studies. While we recognize that these observations do not unambiguously characterize the nature of the CV, we are able to put constraints on the binary parameters and the spectral type of the secondary. We speculate that LSQ172554.8-643839, like other extremely deep-eclipsing CVs, is also a polar.

**OBSERVATIONS**

After the initial discovery, we immediately scheduled follow-up observations with ANDICAM on the SMARTS 1.3-m telescope at Cerro Tololo, which is able to record near-IR and visible-band images simultaneously. Using service-mode observing time allocated to Yale for targets of opportunity, we recorded several hours of continuous light-curve observations in different filter combinations on each of 7 nights from 2010 Apr 26 to May 18. We also made remote light curve observations in the $Q_{st}$-band with the LSQ on 2010 Apr 29, and in user mode at several different epochs with the CCD camera on the SMARTS 1.0-m telescope at Cerro Tololo. Finally, we measured the visual spectrum of the target at 6 different light-curve phases on 2010 Jun 10 using the Goodman High Throughput Spectrograph (GHTS) on the SOAR 4.3-m telescope at Cerro Pachon. Table 1 lists the observing circumstances for each night of the light-curve observations, including the telescope, filter, starting Julian date and time span, exposure time, and total number of observations. Photometric nights are noted. The intervals between observations were 100 s for the LSQ data, 58-73 s for the J-band data, and 250 – 400 s for the other data sets. Table 2 lists the observing circumstances with the GHTS. All J-band observations were simultaneous with the visible-band observations made the same night.

The light-curve observations we made with LSQ have the best time resolution of all our follow-up sequences, and the longest continuous coverage. No effort was made to calibrate to photometric standards. The data were used only to determine the brightness changes relative to a nearby and brighter field star. The IRAF "phot" routine was used, with the flux aperture fixed at a 1.5-pixel radius, comparable to the seeing in the images. Magnitude errors were determined from the statistical noise in the target flux and the nearby background.

With the SMARTS 1.3m telescope, we initially dedicated one night to simultaneous J/R observations. Later, we observed in J/V and J/B on alternate nights. Observing conditions were photometric on several of these nights. We were able to calibrate the target magnitudes in each band to stellar standards. Our method was to use observations of Landolt stars taken on the photometric nights to calibrate the B, V, and J magnitudes of selected field stars in the target images also taken on the photometric nights. These fields stars were then used as secondary standards to calibrate the target magnitudes in every target image, including the data we recorded on non-photometric nights. A detailed description of the method appears in Rabinowitz, Schaefer, & Tourtellotte (2007).

With the SMARTS 1.0m telescope, we observed on different nights in Gunn z and Johnson U and I. Unfortunately, none of these nights were photometric, and we were unable to calibrate selected field stars for absolute photometry. We used these data only for relative photometry (as noted in Table 1), measuring the magnitudes by the same method we used to analyze the LSQ data (except that several nearby field star were used for relative photometry, instead of just one) .

Our spectroscopic observations consist of an initial series of three 5-min exposures timed to cover the period immediately preceding the deep eclipse, and three 10-min exposures to cover the period immediately following the deep eclipse. We employed the 300 l/mm grating, a 1.03" slit, and a 2x2 readout binning. This provided a wavelength coverage of 3550 - 8870 Å, an instrumental resolution of 2.6 Å/pix and a spectral resolution of about 10 Å as measured from the FWHM of comparison lamp lines. We aligned the slit with the parallactic angle at the start of the observations. We used standard IRAF routines to process the data in order to obtain fully-calibrated spectra. After sky subtraction and extraction of the one-dimensional spectrum, we proceeded with wavelength and flux calibrations. For the latter, we employed observations of the spectrophotometric standard star LTT 4364 obtained during the same night with a wide (3.0") slit. We also corrected the spectra for continuum atmospheric extinction and telluric absorption bands. For the telluric correction we used a high signal-to-noise ratio spectrum of LTT 4364 obtained using the same slit as for the science spectra.

1. RESULTS

*3.1 Eclipse ephemeris*

To compare all our observations, which were recorded over a time span of ~1 month, we corrected all exposure times to the reference frame of the solar system's barycenter. An initial determination of P was then determined from the LSQ data alone. A much more accurate measurement was then obtained by searching for the period value near this initial determination that best predicts the central times of the deep eclipses we observed in all the light curves over the one-month span of our observations (18 separate epochs). The resulting ephemeris for the barycentric Julian date, $T_o$, at central eclipse versus cycle count, E, is:

$$T_o = BJD\ 2455312.5995 \pm 0.0003\ + P\ E \qquad (1)$$

where $P = 0.065734 \pm 0.000001$ days (94.657 min).

*3.2 $Q_{st}$-band Light curve Shape*

Figure 2 shows the $Q_{st}$-band light curve, $Q_{st}^*$ versus θ, where $Q_{st}^*$ is the magnitude of the target relative to a nearby field star and θ is the orbital phase. The zero point for the magnitude scale has been adjusted to match the peak $Q_{st}^*$ value to the peak V value observed later with the SMARTS 1.3m. More than four periods have been folded to yield the result. To display a full cycle clearly, we over-plot the same data shifted to the right one whole phase. Figure 3 shows an expanded plot of the values near θ=1. Note that all the $Q_{st}^*$ values in the middle of the deep eclipse, from θ = 0.98 to 1.03, are 1-σ upper limits to $Q_{st}^*$. The signal from the target at these phases is not significantly different from the sky background.

The resulting light curve consists of three contiguous regions with distinctly different behavior:

Region 1: A deep and narrow eclipse (width $\Delta\theta \sim 0.05$ centered at $\theta = 0$), with duration ~4.7 min, where there is no detectable signal from the source (see Sec 4.2 for a detailed discussion of the uncertainty for $\Delta\theta$).

Region 2: A wide and relatively shallow eclipse ($\theta = 0.03$ to $0.45$), with duration ~39.8 min, and where the magnitude brightens linearly with time by 0.5 mag, starting at a value 2 mag fainter than peak.

Region 3: The un-eclipsed part of the light curve ($\theta = 0.45$ to $0.98$), with duration ~50.2 min, where the signal reaches peak brightness, and varies roughly as a sinusoid with peak-to-peak amplitude ~1.0 mag and frequency nearly double that of the whole light curve.

*3.3 $Q_{st}$-band Eclipse Depth*

To determine a lower limit to the depth of the $Q_{st}$-band eclipse, $\Delta Q_{st}$, we averaged the ten LSQ images with phases closest to middle of the deep eclipse (after subtracting the sky background) to produce a deep image, $I_{deep}$. With no source visible above background in $I_{deep}$, we measured the rms signal, $\sigma$, of the background and calculated the limiting flux, $F_{lim} = 3\sigma\pi^{1/2}$, that we would expect for a target with brightness 1-sigma above background inside an aperture of 3-pixel radius. Using the same radius aperture, we also measured the flux, $F_o$, of the target in an image, $I_o$, recorded at $\theta = -0.05$. We also measure the flux for a nearby field star in both $I_{deep}$ and $I_o$. Normalizing $F_{lim}$ and $F_o$ to the field star flux, we find the 1-sigma limiting flux in $I_{deep}$ to be 195 times fainter than target flux at $\theta = -0.05$. Hence, we obtain $\Delta Q_{st} = 5.7$ mag.

*3.4 Light Curve Wavelength Dependence and Phase-Dependent Color*

Figure 4 shows the phased light curves we obtained from our SMARTS 1.3m and 1.0m observations in different wavelength bands. Here we have assigned arbitrary magnitude zero points to those light curves that we did not calibrate to standards (the y-axis label appears with an asterisk in these cases). For the other cases, the magnitude zero points are calibrated to standards as described above. All light curves are repeated with a horizontal shift of one whole period. The vertical scale is identical for each band. In all bands except J, a deep eclipse was detected where the source became undetectable. The eclipse appears as a gap in the light curves at phase $\theta=1$. To guide the eye, we draw a line connecting the points for each light curve, but only for the phases from $\theta = 0$ to 1. We leave out the connecting line from $\theta = 1$ to 2.

Inspection of Fig. 4 shows that the light-curve shape is generally repeated in each of the visible bands (U to z), with regions 1, 2, and 3 evident, as defined above. The most significant change is an increase in the slope of region 2 as the band wavelength changes from U to z. There is also a slight change in the shape of region 3 from sinusoidal to saw tooth. In the J-band there is a dramatic change. The deep eclipse is completely absent and there is no clear distinction in the light curve shape in the phase ranges corresponding to regions 2 and 3. Interestingly, there is a sudden drop in brightness by ~1 mag at θ=0.9. This precedes the onset of the deep eclipse that is apparent in the optical bands by ~5 minutes. During the phase range covering the deep eclipse, the J band brightness is increasing.

The night-to-night variations in the shape of the light curve are apparent in the scatter of the V-band observations, which span 7 nights. The scatter ranges from +/- 0.1 mag for region 2 to +/- 0.5 mag at the peak of the light curve where region 3 begins. There do not appear to be any long-term changes. The $Q_{st}$-band light curve recorded Apr 29 is nearly identical to the B- and V-band light curve observed May 10-18. Hence, there is no evidence of any outbursts or significant change in activity level over a 3-week interval.

Figure 5 show the band ratios (in magnitudes) with respect to V as a function of orbital phase. These we obtain by subtracting each measured magnitude by the V-band magnitude at the nearest orbital phase. For reference, the top panel shows the phased V-band light curve from Fig. 4. Note that the zero point of the magnitude scale for U*-V, V-I*, and V-z* is arbitrary since we do not have absolute photometry for the U, I, and z bands. In each panel, we omit magnitude ratios measured within and near the deep eclipse because the optical flux is too faint to measure accurately, or not detectable.

Inspection of Fig. 5 shows that the B-V and V-R are relatively constant with orbital phase. Although there is significant variation, it is generally less than 1 mag peak to peak. In Table 3, we show the phase-averaged values for B-V and V-R for regions 2 and 3 of the light curve. The source is slightly redder in region 3 than in region 2 (the mean B-V increases by 0.23 mag) but V-R only increases by 0.09 mag. Table 3 also lists a lower-limit, V-J > 7.1, that we estimate for region 1. This is the V-band lower bound ($V_{lim}$ = 24.1) we estimate at the center of the deep eclipse minus the phase-averaged J-band magnitude (J = 17.1 mag) we measure from θ = -0.02 to 0.03. We derive $V_{lim}$ from the magnitude, V = 18.4, just before the onset of the eclipse (at θ = -0.05) plus $\Delta Q_{st}$ derived above. The $Q_{st}$- and V-band light curves are likely to have the same eclipse depths because they are centered on the same wavelength (5500 Å).

*3.5 Spectra*

Figure 6 shows the six different spectra we recorded for the binary, with the first three spectra (starting at the bottom) corresponding to phase region 3, and next three spectra corresponding to phase region 2. Each panel is labeled with orbital phase, $\theta_{mid}$, corresponding to the mid-time of the exposure (see also Table 2), and determined using Eq. 1.

It is immediately apparent that Balmer-series emission lines are present at all observed phases, with $H_\alpha$ (6563 Å), $H_\beta$ (4861 Å), $H_\gamma$ (4341 Å), $H_\delta$ (4102 Å) and $H_\epsilon$ (3970 Å) clearly resolved in phase region 2 (θ = 1.08, 1.19, and 1.30). In this region, the spectra also clearly resolve He I (5876 Å). It also apparent that the continuum varies significantly with orbital phase, with a significant wavelength-independent brightening in region 3 ((θ = 0.72, 0.78, and 0.84) relative to region 2. This is consistent with the phases of maximum intensity in the optical light curves. The continuum is noticeably flat at all wavelengths in region 3, but slopes towards blue wavelengths in region 2.

Figure 7 shows the same spectra as Figure 6, expanded around the $H_\alpha$ line. It is clear that the shape, width, and center of the lines are changing with orbital phase owing to phase-dependent variation in the source region's radial velocity. Similar variations also appear for the $H_\beta$ and $H_\gamma$ lines. Using fitting routines in IRAF ("splot" with Voigt fit), we have measured the line centers (shown by the vertical dashed lines in Fig. 7) and widths versus phase for all three emission lines at each light-curve phase. For each emission line, Fig. 8 shows the resulting radial velocity values as a function of $\theta$. Assuming a source in synchronous rotation, we fit a best-fit cosine function for each line of the form

$$v_r = \gamma + K*\cos(2\pi[\theta - \theta_0]) \qquad (2)$$

where $\gamma$ is the mean radial velocity, K is the amplitude of variation, and $\theta_0$ is the phase at which $v_r$ reaches a maximum (largest red shift). Table 4 lists the parameters of resulting fits, which yields rms residuals of 16, 26, and 36 km s$^{-1}$ for $H_\alpha$, $H_\beta$, and $H_\gamma$, respectively (about ¼ pixel resolution). For each emission line, the errors on the fitted parameters are calculated assuming an observational uncertainty in $v_r$ equal to the rms with respect to the best fit, and taking the ± 1-sigma error for each parameter to be the range that keeps $\chi^2$ less than the minimal $\chi^2$ + 1. Table 4 also shows the mean width (FWHM) measured for each line for the Lorentzian component of the Voigt fit, which dominates in nearly all cases.

## 4. DISCUSSION

The observations we present above are not sufficient to uniquely characterize the nature of our target. Unambiguous characterization would require polarization measurements, long-term monitoring, photometry with higher time resolution, observations in ultraviolet and x-ray wavelengths, and a more extensive spectral series. In the following discussion, we attempt to put some preliminary constraints on the binary parameters, spectral type of the secondary, and we offer some speculation regarding the source of activity.

*4.1 Emission Regions*

We begin with the hypothesis that the source is an eclipsing polar, like J015543.40+002807.2. Our observations of a very short-period light curve and two eclipses – one very narrow and deep, the other relatively shallow and persisting for one half rotation – match the phenomena we would expect for such a system. In our case, there is at least one bright spot on the surface of the WD in synchronous rotation with the orbit. The spot rotates behind the WD each half period, producing the wide, shallow eclipse (region 2). Just after it rotates from view, there is a brief eclipse by a cool companion star, invisible at optical wavelengths. This produces the short, deep eclipse. The average color ratios we measure for region 2 (B-V = -0.02 and V-R = 0.20), when the hot spot is presumably occluded but the WD is visible, are consistent with a typical WD (Schwartz 1972). Also, the spectral shape and Balmer emission lines we observe at phases covering region 2 ($\theta$ = 1.08, 1.19, and 1.30) are similar to spectral features observed by Schmidt et al (2005) for polar J015543.40+002807.2.

There are some features that we expect for a polar, but do not see. Cyclotron emission should produce broad humps in the red part of the optical spectrum. These may be present at a low level in our data, or they may not be present at the orbital phases we observed. Schmidt et al (2008) find the intensity of cyclotron emission varies with time and orbital phase for polar SDSS J0921+2038. Polars are also known to produce strong He II emission at 4686 Å. While this line appears in our spectra at $\theta$ = 1.08, 1.19, and 1.30, it is much weaker than typically observed. However, the spectrum of SDSS J0921+2038 also shows He II at a weak level, comparable to the level we see in our candidate polar. For polar UW Pic, Romero-Colmenero et al. (2003) observer the He II line to vary from weak to strong on yearly time scales. Hence, the absence of cyclotron emission and of strong He II lines in our limited spectral series does not rule out a polar.

A second possibility is that we are observing a system with an accretion disc and a hot spot where the stream from the secondary impacts the disc edge. The increase in source intensity just prior to the deep eclipse ($\theta$ = 0.6 to 0.95) might then be explained as an orbital hump, as in Z Cha (W95) or IY UMa (Patterson et al. 2000). However, this would not explain the extremely sharp rise in optical intensity we observe at $\theta$ = 0.45, where region 2 ends. From orbital phase 0.452 to 0.464 the source brightens by 0.9 magnitudes. This is much faster than the typical brightness change expected for the egress of hot spot from behind a disk. For example, the orbital hump in Z Cha requires 20% of an orbital period to reappear. In IY UMa, the hump rises in 30% of a period. On the other hand, Schwope & Mengel (1997) observed the eclipsing polar EP Dra to brighten in the V band by ~50% over very narrow range of orbital phase, outside of eclipse (0.72 – 0.75).

The rapid intensity rise we see at the end of region 2 is peculiar, and may be consistent with neither scenario discussed above. There are other unexplained features in our light curve: the drop in optical brightness from $\theta$ = 0.6 to 0.8, just after the steep increase at the end of region 2; also the sharp drop in the J-band brightness at $\theta$ = 0.95, just before the onset of the deep optical eclipse. Perhaps a complete model could explain these features, taking into account the intensity versus emission angle for the hot spot, the inclusion of an accretion disk in the system, and eclipses of the WD by the accretion column. However, such detailed modeling is beyond the scope of this paper and should wait for more extensive follow-up observations.

*4.2 Binary Parameters*

To estimate the binary parameters – mass and radius of the secondary ($R_{sec}$ and $M_{sec}$), the binary separation, a, and the inclination, i – we assume the secondary has relaxed to a circular orbit and fills its Roche lobe. The strong Hydrogen emission lines we see in the spectrum are evidence that there is infalling gas, and hence a Roche-lobe filling secondary. Following Patterson et al. (2005) and Knigge (2006, hereafter K06), we also assume a WD mass, $M_{WD}$ = 0.75 (here and in the following discussion, we use the solar values for radius and mass units). This value for $M_{WD}$ matches the values determined for other eclipsing CVs to within 20%. We also assume a WD radius, $R_{WD}$ = 0.013, which matches the radii measured for WDs with masses 0.50 to 0.90 to within +/- 0.002 (Parsons et al. 2010).

With these assumptions, we can now use the empirical mass/radius relation for CV secondaries determined by K06, based on an analysis of super-humping in eclipsing systems:

$$R_{sec} = A(M_{sec}/M_o)^\alpha \qquad (3)$$

where $A = 0.110 \pm 0.005$, $M_o = 0.063 \pm 0.009$, $\alpha = 0.21(+0.05,-0.10)$ for $M_{sec} < 0.063$ and $A = 0.230 \pm 0.008$, $M_o = 0.20 \pm 0.02$, $\alpha = 0.64 \pm 0.02$ for $0.063 < M_{sec} < 0.20$ and $P < 2.15$ h. Here the change in the constants at mass 0.063 relates to the evolutionary status of the secondary. Owing to the competing effects of mass transfer and to the loss of angular momentum by magnetic breaking and gravitational radiation, CVs are expected to evolve from longer to shorter periods with the secondary continuously losing mass. Eventually the secondary becomes too light for hydrogen burning and remains supported instead by electron degeneracy pressure. Thereafter, with further mass loss, the secondary grows in size and P also increases. Hence, for $M_{sec} < 0.06$, Eq. (3) describes secondaries that have evolved through a period minimum (post-bounce systems). For $0.063 < M_{sec} < 0.20$, the equation describes pre-bounce systems. For $M_{sec} > 0.20$, K06 provides a third set of values for the constants in Eq. 3, corresponding to CVs that have not yet evolved below the period gap ($P > 3.18$ h). K06 also warns that the post-bounce mass/radius relation is very uncertain, as it is derived from observations of only a few CVs.

We can also use the density/period relation from W95, valid for Roche-filling secondaries, which K06 restates as

$$R_{sec} = 0.2361[M_{sec}^{1/3} P^{2/3}] \qquad (4)$$

where P is in hours. Combining Eq. 3 and 4, we thereby determine $R_{sec}$ and $M_{sec}$ from P. We also use Kepler's law to determine a from P, $M_{sec}$, and $M_{WD}$.

The final binary parameter, i, is fixed by the eclipse viewing geometry and $\Delta\theta$:

$$\sin(i) = [1-(R_{sec}/a)]^{1/2}/\cos(\pi\Delta\theta) \qquad (5)$$

Here $\Delta\theta = \theta_e - \theta_i$, where $\theta_i$ and $\theta_e$ are the respective orbital phases of mid ingress and mid egress. Because our light curves do not have the time-resolution to reveal these contact points, we estimate lower and upper bounds as follows (indicated by the dashed lines in Fig. 3). The eclipse ends no later than $\theta = 1.037$, where we make our first post-eclipse detections. We thus have $\theta_e < 1.036 - R_{WD}/2\pi a$, where $R_{WD}/2\pi a$ is half the egress phase duration. Egress also begins no earlier than $\theta = 1.022$, where we have our last null detection and the source intensity is within a few percent of minimum. Hence, $\theta_e > 1.022 + R_{WD}/2\pi a$. By similar reasoning, given our last pre-eclipse detection at $\theta = 0.970$ and our first null detection at $\theta = 0.982$, we have $\theta_i > 0.968 + R_{WD}/2\pi a$ and $\theta_i < 0.982 - R_{WD}/2\pi a$. With these extremes for $\theta_e$ and $\theta_i$, we thus have $\Delta\theta > 0.040 + R_{WD}/\pi a$ and $\Delta\theta < 0.067 - R_{WD}/\pi a$.

Solving the above equations for $R_{sec}$, $M_{sec}$, a, and i, we have separate solutions for the pre- and post-bounce cases. The resulting values and their uncertainties are listed in Table 5. The table also lists two values for $\Delta\theta$, corresponding to the two solutions for a. We derive the results by running iterative solutions, assigning Gaussian-distributed random values to A, $M_o$, $\alpha$, $\Delta\theta$, $M_{WD}$, and $R_{WD}$ within their respective uncertainty ranges on each trial. This provides a distribution of solutions for each parameter for which the mean and standard deviation are the value and uncertainty respectively listed in Table 5. Note that Eq. 5 restricts the range of values the parameters can take because the deep eclipse can not occur if $R_{sec}/a > \sin(\pi\Delta\theta)$.

Comparing the two solutions, it is apparent that the post-bounce solution yields a secondary that is much less massive ($M_{sec} = 0.034 \pm 0.012$) and smaller ($R_{sec} = 0.102 \pm 0.011$) than the pre-bounce case ($M_{sec} = 0.104 \pm 0.025$, $R_{sec} = 0.150 \pm 0.012$). The values for a differ by only a few percent, since the WD mass dominates the system in both cases. Both the pre- and post-bounce solutions require a highly inclined orbit, with i > 80 deg. While the pre-bounce mass is typical of known short-period CVs, the post-bounce mass would be extremely unusual. In the Ritter & Kolb catalog, only 2 of the 128 CVs with measured secondary masses have a lower value. However, post-bounce systems are predicted to dominate the CV population because they are a long-lived end state (Kolb 1993). Recently, faint CVs detected in spectroscopic surveys by the Sloan Digital Sky Survey (SDSS) reveal a substantial fraction of post-bounce systems, supporting the existence of the large population (Littlefair et al. 2008, Gänsicke et al. 2009). Given that our source is faint, comparable in magnitude to the CVs detected by SDSS, the likelihood that it could be a post-bounce system is enhanced. In the next section, we consider the possibility we are seeing such a low-mass companion based on photometric constraints.

*4.2 Secondary spectral type and distance*

From our photometry during the deep eclipse, we see that the secondary is a very red object with V-J > 7.1. According to the table of grizJK colors for low-mass stars compiled by Dupuy and Liu (2009), and using the conversion from g-r to g-V from Fukugita et al (1996), this color corresponds to a spectral type M8 or later. This is extremely rare among the known white-dwarf binaries. Only 2 of 153 WD binaries in the 2MASS survey by Hoard et al. (2007) and only one of the 91 CV secondaries tabulated by K06 have such a late spectral type. Mass estimates for M8 stars range from 0.02 to 0.08 (Riddick, Roche, & Lucas 2007). This overlaps our post-bounce value for $M_{sec}$ but is only marginally consistent with the pre-bounce value. Hence, if the very red color we measure is a valid indication of the spectral class of the secondary, it supports the possibility that we are observing a very unusual, low mass companion in a post-bounce system.

We must also consider the possibility that our spectral classification is in error, owing perhaps to a J-band emission from an extended source adding to the emission from the secondary at deep eclipse. However, if the pre-bounce solution for the binary mass were correct, we would expect a much bluer color for the secondary. The fits to spectral type and color versus period provided by K06 yield spectral type M6.3 and V-J = 5.5 at the period we observe for our source. Hence, a pre-bounce solution implies that IR emission contaminates our J-band measurement at mid eclipse by 1.6 mag, or a factor of 4 relative to the secondary. The source would have to be located far from the WD primary and its hot spot, and it would have to be much fainter than these sources at optical wavelength. Otherwise the J-band light curve would mirror the optical light curve. A weakly emitting accretion disk or accretion column might meet these criteria.

Assuming our spectral classification is correct and that the secondary's magnitude-color relation is the same as for isolated low-mass stars, then we can determine the minimum absolute J magnitude, $M_J$, and minimum distance, D, to the binary. From Hawley et al (2002), who fit $M_J$ versus spectral class, we have

$$M_J = 8.83 + 0.29*\text{spectral type}$$

where spectral type = 0 to 10 for M0 to M9, and 11-20 for L0 to L9. For spectral type M8.0 or later, this yields $M_J > 11.2$. Then with J=17.1 at the deep eclipse midpoint, we obtain distance D = 150 parsec.

*4.3 Doppler Constraints*

The fits we derive to the radial velocity curves for $H_\alpha$, $H_\beta$, and $H_\gamma$ yield consistent values for the amplitude, K, and phase, $\theta_o$, of the radial velocity oscillation. Averaging the results, we obtain K = 500 +/- 22 km s$^{-1}$ and $\theta_o$ = -0.01 +/- 0.01. Assuming the Doppler shift is caused by the Keplerian motion of the source region on a circular orbit (such as the hot spot on the edge of an accretion disk), then a maximum redshift at $\theta_o$ = -0.01 places the source region at a phase 90 deg ahead of the secondary, which is not consistent with a typical accretion-disk morphology. Furthermore, the source orbit would have radius, $r_H$, given by

$$r_H = KP / 2\pi\sin(i)$$

or a limiting lower value $r_H > 0.65$ for i=90 deg. However, the maximum radius we can expect for the accretion disk, $r_d*$, is much smaller. W95 provides the following expression (Eq. 2.61):

$$r_d* = 0.6a/(1+M_{sec}/M_{WD})$$

This yields $r_d* < 0.369 \pm .030$ (post-bounce) and $r_d* <0.315 \pm 0.029$ (pre-bounce). Thus it is clear that emission line properties we observe are inconsistent with a hot spot on the edge of an accretion disk.

On the other hand, the large amplitude of the radial velocity oscillations, the near zero phase at which they reach maximum redshift, and the extreme Doppler broadening we observe for the emission lines (600 to 1300 km s$^{-1}$) are consistent with the radiation emitted from the accretion column in a typical polar (W95). In this case, the Doppler shift derives from the infall velocity of the gas pulled from the secondary and falling onto to the WD. Near $\theta = 0$ deg, the view is directly along the accelerated gas flow, and the red shift is maximal. Furthermore, the light-curve variations we observe in region 2, where the hot spot is visible, would then be related to the changing orientation of the accretion column with respect to the line of sight, as for V834 Centauri (W95).

## 5. CONCLUSIONS

We have presented initial follow-up observations of an intriguing new cataclysmic variable and set constraints on the binary parameters, secondary spectral type, and source of the emitting region. Although the picture is sketchy, we believe we can best explain the observations if the system is a polar, consisting of a WD in synchronous rotation and a near-surface hot spot that is occluded by the WD at nearly the same time that the WD is eclipsed by the secondary. This model explains the extended flat part of the light curve (region 2) after the deep eclipse ends, the fast rise in brightness at the end of region 2, the extreme Doppler broadening of the emission lines, and the large amplitude of the radial velocity oscillations. The conclusion is uncertain, however, because we do not see cyclotron humps or strong He II lines in the spectrum. On the other hand, the phasing and large amplitude of the radial-velocity oscillations we observe are not consistent with emission from a hot spot on an accretion disk. A definitive confirmation will require the observation of polarized emission characteristic of polars. Much higher time-resolution photometry will also be required to clearly resolve features of the deep eclipse ingress and egress which are diagnostic CV morphology. X-ray or far UV follow-up observations with XMM might also determine the nature of the source depending on its level of activity.

It is intriguing that our observations suggest a secondary of very late spectral type. LSQ172554.8-643839 could be a rare example of CV with a very low-mass, degenerate dwarf companion, which are valuable for testing evolutionary models of CVs beyond the period bounce, and for constraining mass-radius relation for low-mass stars. This is unlikely, given the rarity of such discoveries, but IR observations with the Spitzer telescope would reveal the emission from the secondary if is it is a degenerate dwarf. Since we discovered this object incidentally, by inspection of a tiny fraction of the survey data acquired to date, it is likely that many more eclipsing CVs could be readily discovered with LSQ, including more interesting systems like LSQ172554.8-643839.

ACKNOWLEDGEMENTS

This work was supported by the U.S. Dept. of Energy and The National Aeronautics and Space Administration. P. R. and S. H. acknowledge partial support from the Center of Excellence in Astrophysics and Associated Technologies (PFB-06). G.F. acknowledges partial support from the Millennium Center for Supernova Science through grant P06-045-F funded by Programa Bicentenario de Ciencia y Tecnología de CONICYT and Programa Iniciativa Científica Milenio de MIDEPLAN. We especially thank the staff of the European Southern Observatory at La Silla for supporting LSQ, and B. Warner and an anonymous referee for helpful comments and guidance.

**Figure Captions**

Figure 1. Eclipse depth versus peak magnitude for known eclipsing compact binaries compiled by Ritter and Kolb (2003). The star-shaped symbol shows LSQ172554.8-643839.

Figure 2. Relative $Q_{st}$-band brightness versus orbital phase, $\theta$, from La Silla – QUEST observations. The magnitude scale has been adjusted so that the maximum $Q_{st}*$ value matches the maximum V-band brightness in Fig. 4 .

Figure 3. Same as Fig 2, but expanded around phase of deep eclipse. The vertical dashed lines show the range of uncertainty for mid ingress and mid egress.

Figure 4. Brightness in U*, B, V, R, I*, z*, J, and $Q_{st}*$ bands versus $\theta$ (asterisk indicates relative magnitudes only, with no absolute calibration)

Figure 5. Color ratios U-V*, B-V, V-R, V-I*, V-z*, V-J, and V versus $\theta$.

Figure 6. GHTS spectra recorded at phases $\theta$ = 0.72, 0.78, 0.84, 1.08, 1.19, and 1.30. Flux units are erg cm$^{-2}$ s$^{-1}$ Å$^{-1}$.

Figure 7. Same as Fig. 6, but expanded around the $H_\alpha$ line.

Figure 8. Radial velocity determined from $H_\alpha$, $H_\beta$, and $H_\gamma$ at $\theta$ = 0.72, 0.78, 0.84, 1.08, 1.19, and 1.30. Fits are shown as solid ($H_\alpha$), short-dashed ($H_\beta$), and long-dashed ($H_\gamma$) lines.

**Table 1**
Photometry Observing Circumstances

| Telescope | Band | Date | MJD start | Span (h) | Exp (s) | N | Notes |
|---|---|---|---|---|---|---|---|
| SMARTS 1.3m | J | 2010 Apr 26 | 55312.12404 | 1.4 | 36 | 90 | phot |
| SMARTS 1.3m | R | 2010 Apr 26 | 55312.12525 | 1.4 | 240 | 18 | phot |
| ESO 1.0-m | $Q_{st}$* | 2010 Apr 29 | 55315.15427 | 6.6 | 60 | 234 | |
| SMARTS 1.3m | J | 2010 May 10 | 55326.09497 | 2.2 | 36 | 110 | phot |
| SMARTS 1.3m | B | 2010 May 10 | 55326.09619 | 2.2 | 240 | 22 | phot |
| SMARTS 1.0m | I* | 2010 May 10 | 55326.29607 | 3.2 | 300 | 33 | |
| SMARTS 1.3m | J | 2010 May 11 | 55327.19457 | 1.8 | 36 | 100 | phot |
| SMARTS 1.3m | V | 2010 May 11 | 55327.19579 | 1.7 | 240 | 20 | phot |
| SMARTS 1.0m | V* | 2010 May 11 | 55327.25800 | 4.1 | 300 | 36 | |
| SMARTS 1.3m | J | 2010 May 12 | 55328.14954 | 1.8 | 36 | 100 | |
| SMARTS 1.3m | B | 2010 May 12 | 55328.15076 | 1.7 | 240 | 20 | |
| SMARTS 1.0m | V* | 2010 May 12 | 55328.35294 | 2.0 | 300 | 20 | |
| SMARTS 1.3m | J | 2010 May 13 | 55329.27306 | 0.8 | 36 | 50 | |
| SMARTS 1.3m | V | 2010 May 13 | 55329.27428 | 0.7 | 240 | 10 | |
| SMARTS 1.3m | J | 2010 May 17 | 55334.15671 | 1.8 | 36 | 100 | |
| SMARTS 1.3m | B | 2010 May 17 | 55334.15793 | 1.8 | 240 | 20 | |
| SMARTS 1.3m | J | 2010 May 18 | 55335.16961 | 1.6 | 36 | 100 | phot |
| SMARTS 1.3m | V | 2010 May 18 | 55335.17083 | 1.6 | 240 | 20 | phot |
| SMARTS 1.0m | z* | 2010 May 27 | 55343.24783 | 4.7 | 300 | 52 | |
| SMARTS 1.0m | U* | 2010 May 27 | 55343.04094 | 4.5 | 300 | 42 | |

*nights yielding relative photometry only

**Table 2**
Spectroscopy Observing Circumstances

| Date | MJD start | Expt (s) | $\theta_{mid}$ |
|---|---|---|---|
| 2010 Jun 10 | 5357.045648 | 300 | 0.717 |
| 2010 Jun 10 | 5357.049873 | 300 | 0.781 |
| 2010 Jun 10 | 5357.053600 | 300 | 0.838 |
| 2010 Jun 10 | 5357.067639 | 600 | 1.078 |
| 2010 Jun 10 | 5357.075127 | 600 | 1.192 |
| 2010 Jun 10 | 5357.082326 | 600 | 1.301 |

**Table 3**
Phase-Average colors (mag)

|  | Region 2 | | Region 3 | | Region 1 |
| --- | --- | --- | --- | --- | --- |
|  | mean | rms | mean | rms |  |
| B-V | -0.02 | 0.17 | 0.25 | 0.18 |  |
| V-R | 0.20 | 0.15 | 0.29 | 0.15 |  |
| V-J | 2.98 | 0.31 | 1.76 | 0.50 | > 7.1 |

**Table 4**
Best Fit Parameters to Radial Velocity Curves and Emission Line Widths

| Emission Line | $H_\alpha$ | $H_\beta$ | $H_\gamma$ |
| --- | --- | --- | --- |
| $\theta_0$ | -0.03 ± 0.02 | 0.01 ± 0.03 | -0.02 ± 0.02 |
| K (km s$^{-1}$) | 503 ± 40 | 485 ± 34 | 513 ± 41 |
| $\gamma$ (km s$^{-1}$) | -5.7 ± 13 | 62.4 ± 10.8 | 64.5 ± 14.2 |
| rms residual (km s$^{-1}$) | 33 | 26 | 35 |
| mean width (km s$^{-1}$) | 608 ± 93 | 515 ± 266 | 1275 ± 531 |

**Table 5**
Binary Parameters

| Period (d) | $\Delta Q_{st}$ (mag) | $\Delta\theta$ (phase) | $R_{sec}/R_{sun}$ | $M_{sec}/M_{sun}$ | $a/R_{sun}$ | i (°) |
| --- | --- | --- | --- | --- | --- | --- |
| 0.065734 ±0.000001 | >5.7 | 0.053±0.007 | 0.150±0.012 | 0.104±0.025 | 0.648±0.039 | 80.8±2.3 |
|  |  | 0.048±0.006 | 0.102±0.011 | 0.034±0.012 | 0.612±0.045 | 85.9±1.9 |

Notes. The two values listed each for $\Delta\theta$, $R_{sec}$, $M_{sec}$, a, and i correspond to pre- and post-period bounce solutions for the binary parameters, respectively. See text for details.

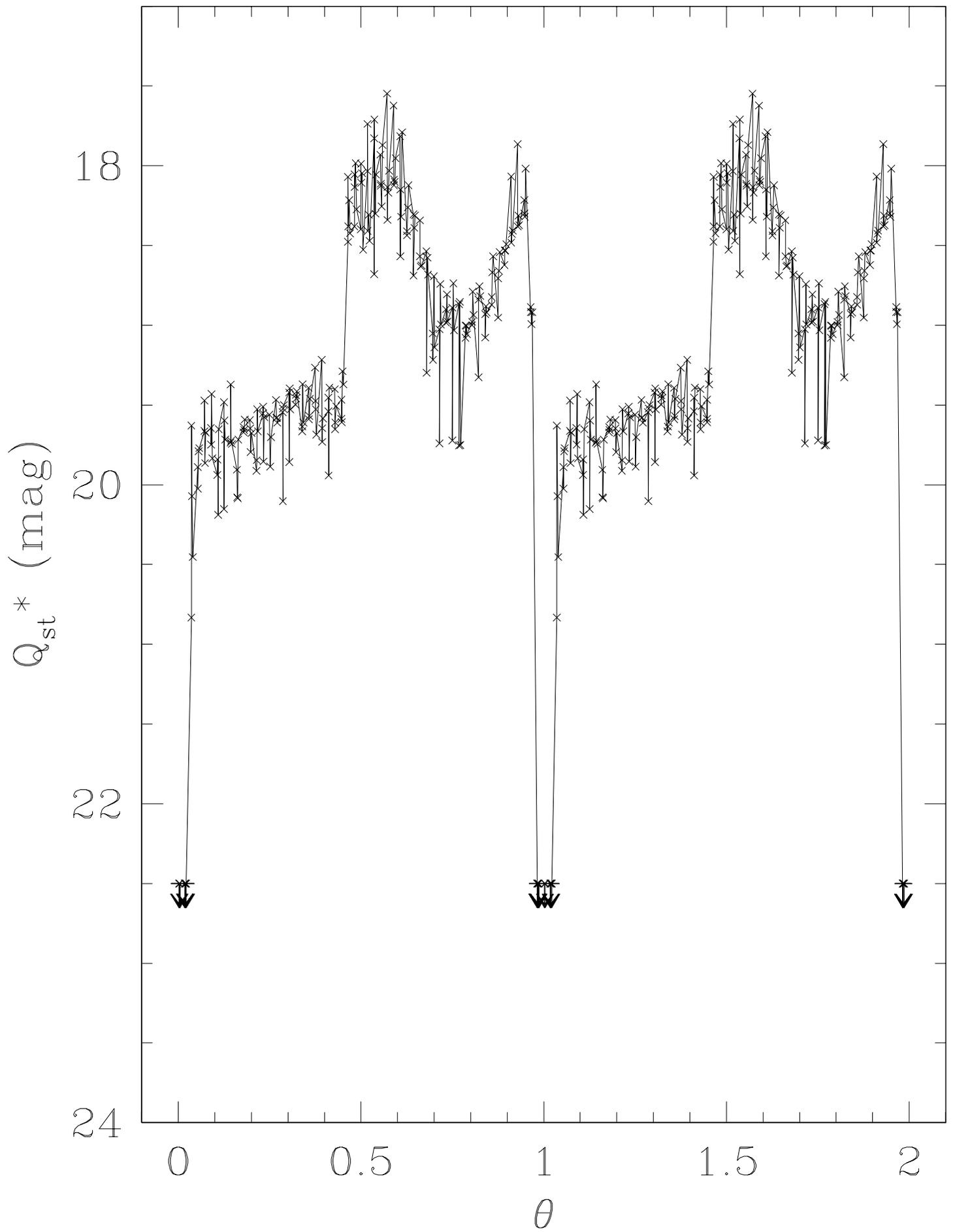

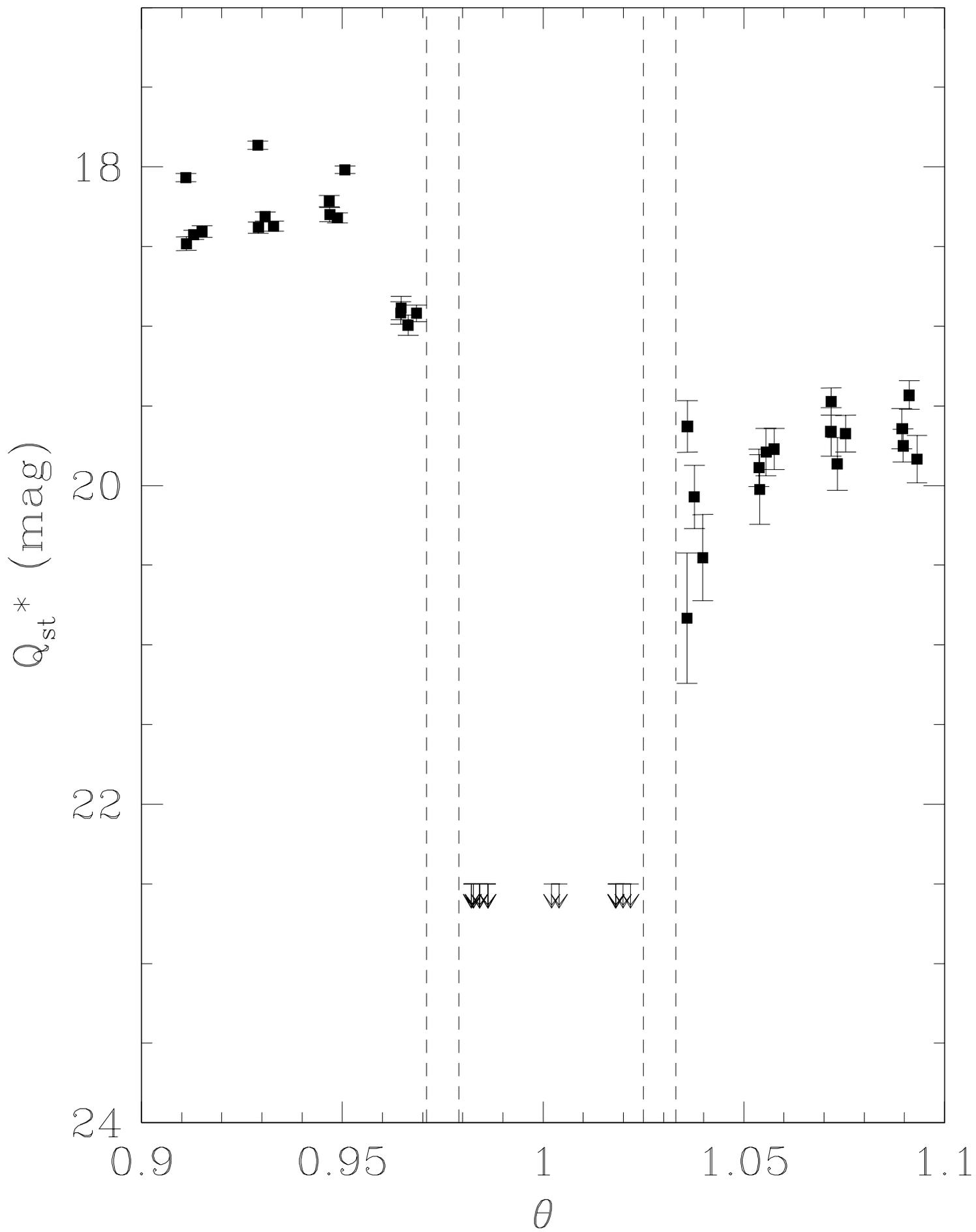

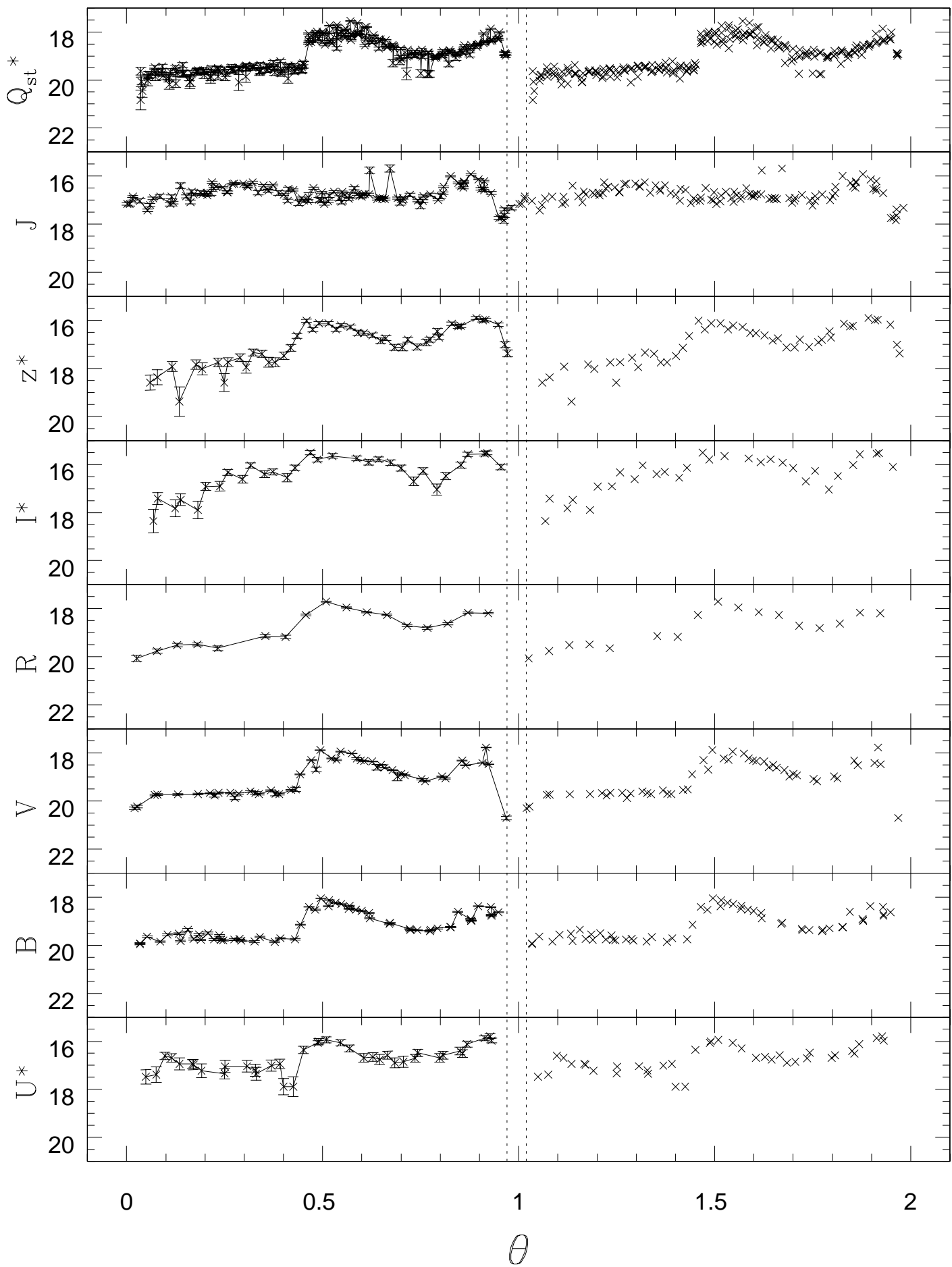

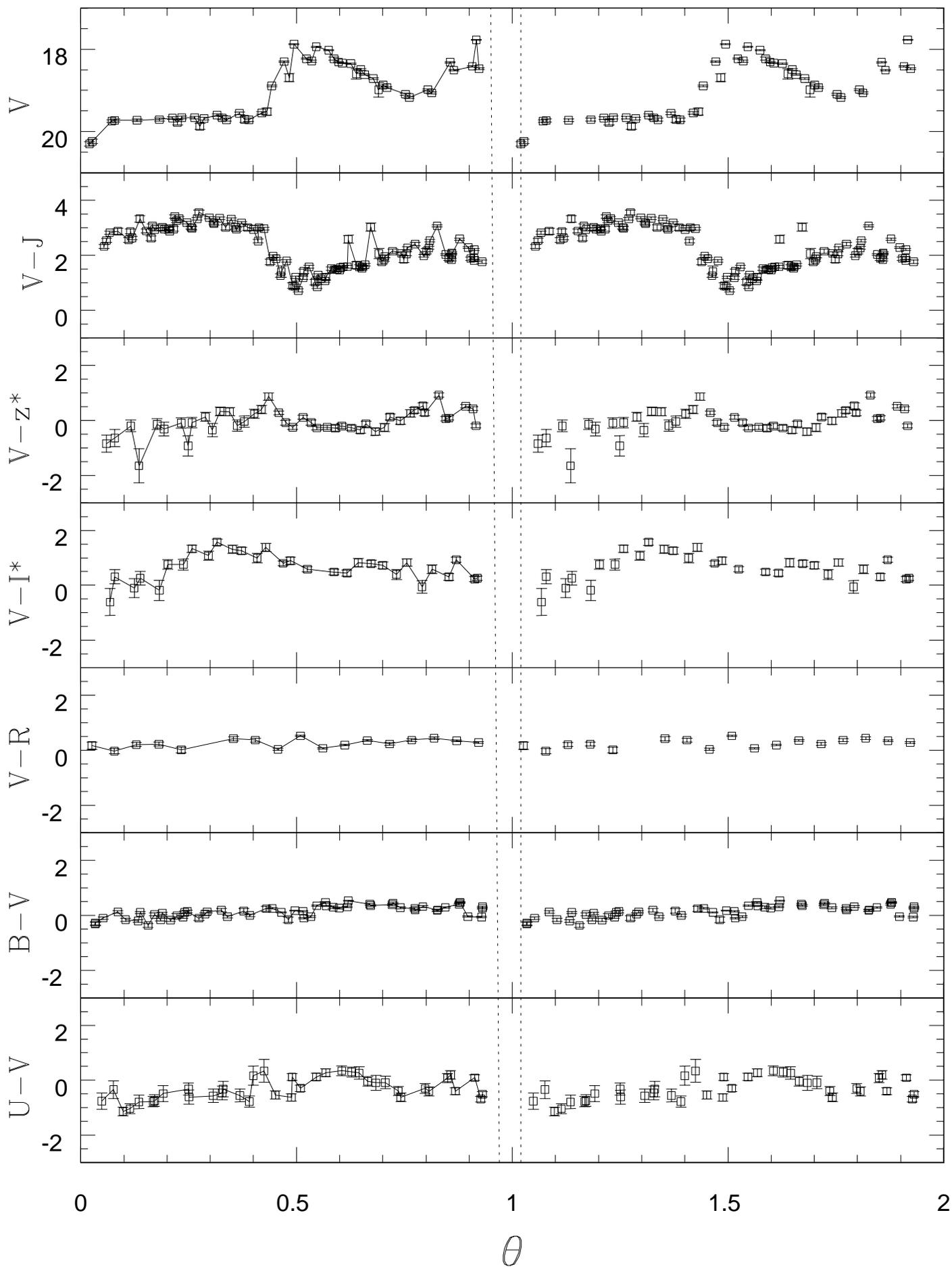

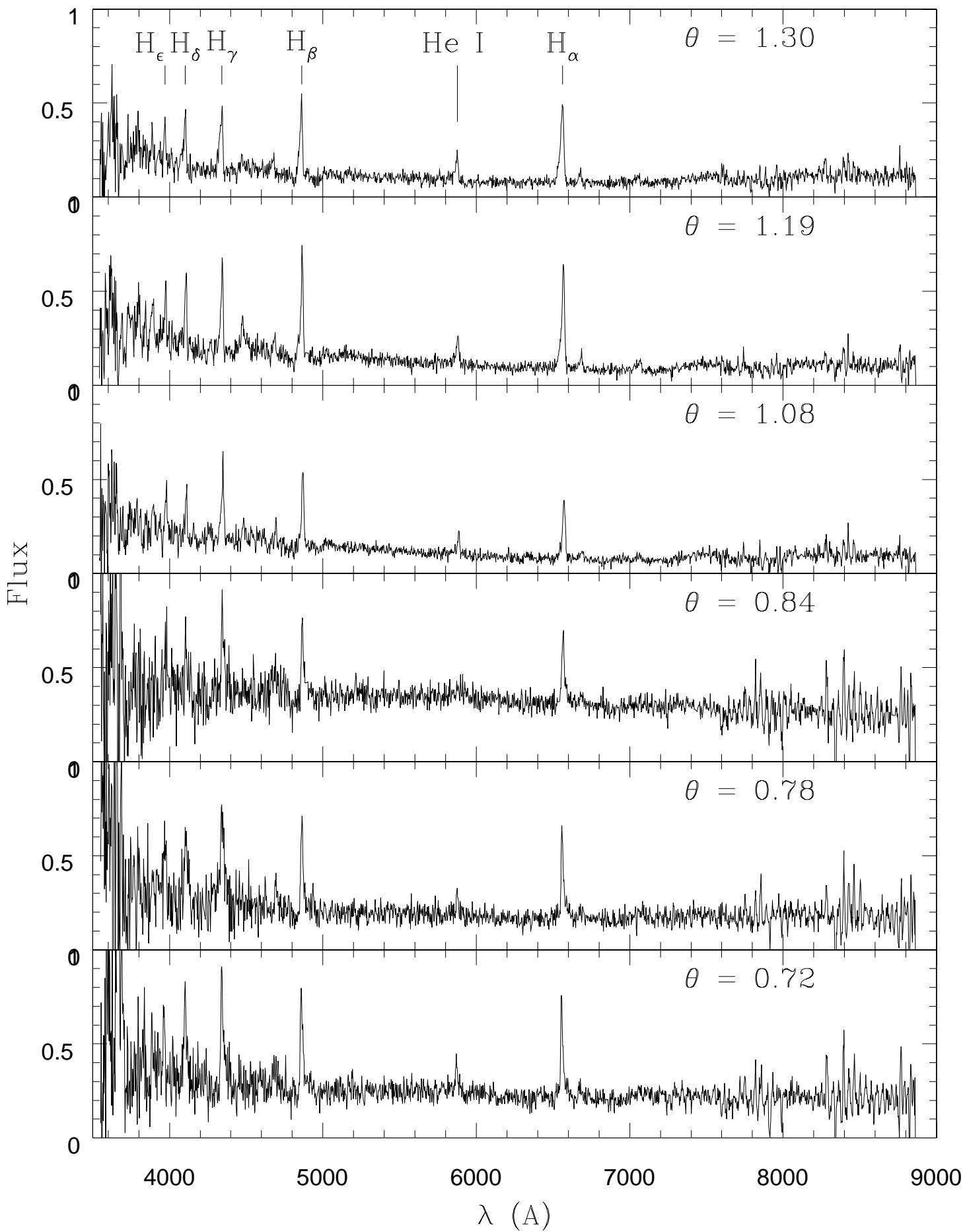

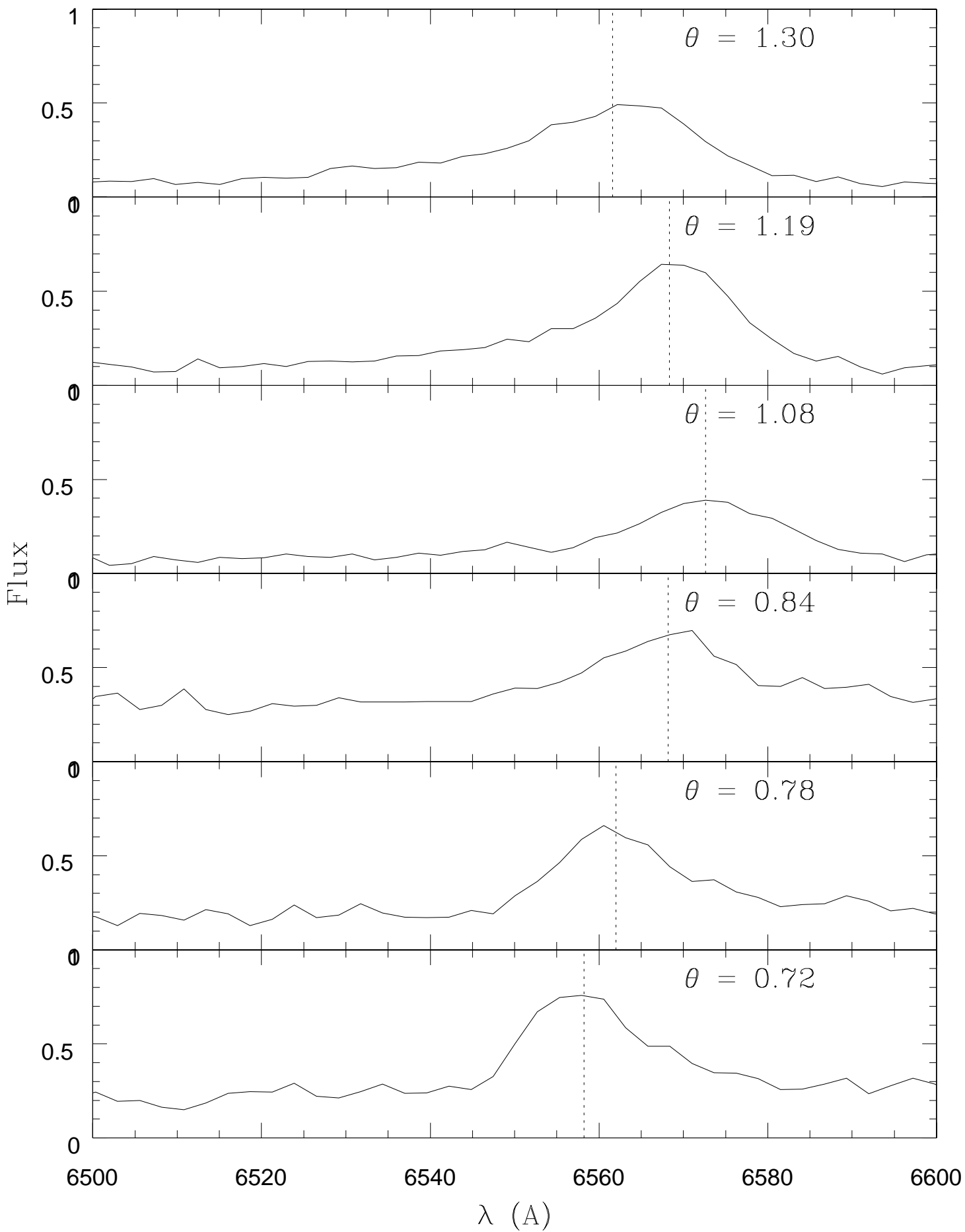

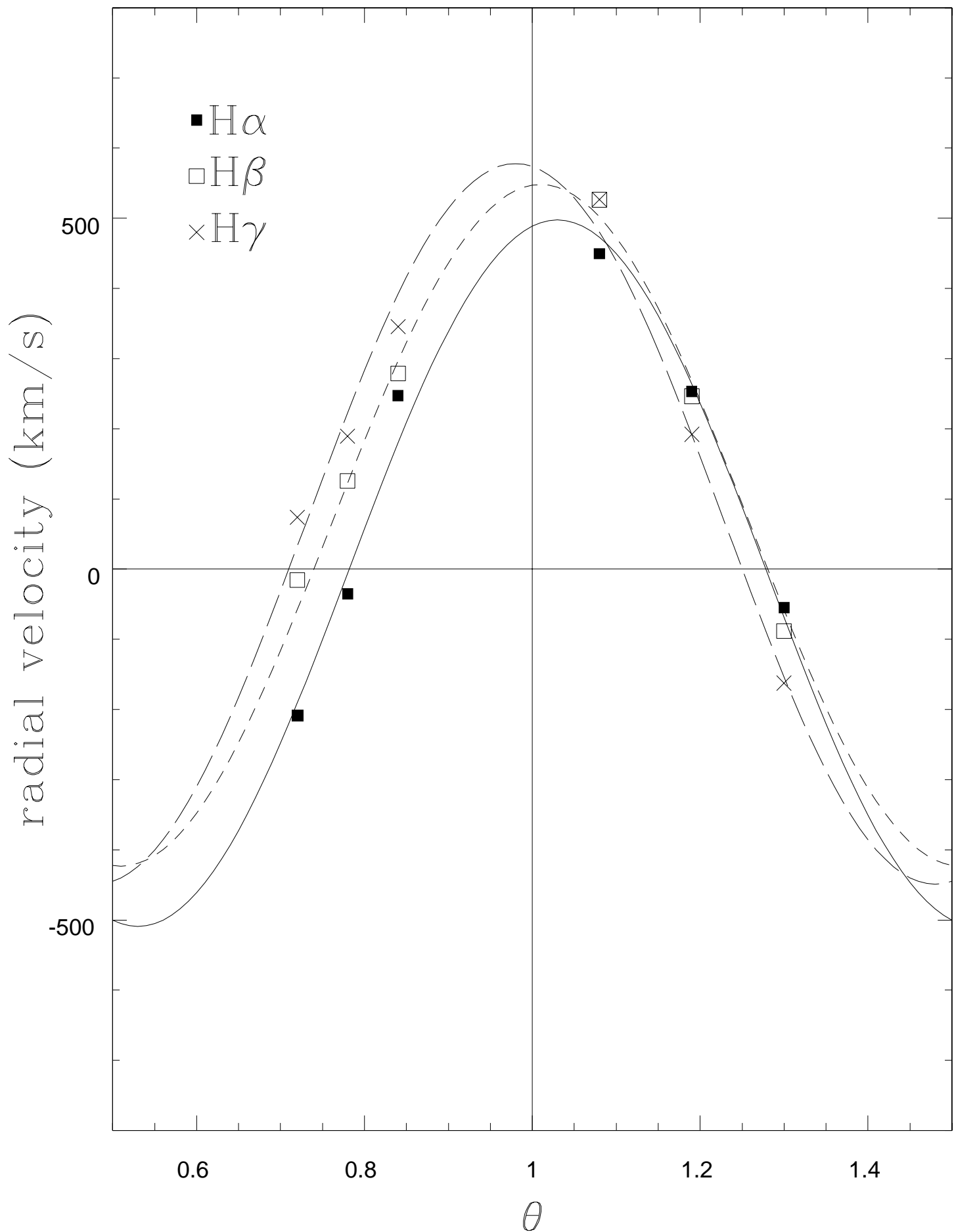